\documentclass{article}
\usepackage{spconf,amsmath,graphicx}
\usepackage{amssymb}

\usepackage{adjustbox}
\usepackage{multirow}

\usepackage{url}
\usepackage{amssymb}

\usepackage{array}
\usepackage{booktabs}

\title{DistillW2V2: A Small and Streaming Wav2vec 2.0 Based ASR Model}
\name{Yanzhe Fu$^\dag$\thanks{$^\dag$equal contribution}, Yueteng Kang$^\dag$, Songjun Cao, Long Ma}
\address{Tencent Cloud Xiaowei, Beijing, China}

\begin{document}
%
\maketitle
\begin{abstract}
Wav2vec 2.0 (W2V2) has shown impressive performance in automatic speech recognition (ASR). However, the large model size and the non-streaming architecture make it hard to be used under low-resource or streaming scenarios. In this work, we propose a two-stage knowledge distillation method to solve these two problems: the first step is to make the big and non-streaming teacher model smaller, and the second step is to make it streaming. Specially, we adopt the MSE loss for the distillation of hidden layers and the modified LF-MMI loss for the distillation of the prediction layer. Experiments are conducted on Gigaspeech, Librispeech, and an in-house dataset. The results show that the distilled student model (DistillW2V2) we finally get is 8x faster and 12x smaller than the original teacher model. For the 480ms latency setup, the DistillW2V2's relative word error rate (WER) degradation varies from 9\% to 23.4\% on test sets, which reveals a promising way to extend the W2V2's application scope.
\end{abstract}
\section{Introduction}
\label{sec:intro}

Self-supervised learning(SSL) achieves great success in ASR. The model is ﬁrst pre-trained on a vast amount of unlabelled data to learn speech representations, and then ﬁne-tuned with little labeled data \cite{baevski2020wav2vec, chung2021w2v, chen2021wavlm, 9688009}. However, the superior performance of the model under the SSL framework typically relies on hundreds of millions or even billions of parameters. Moreover, it can not be used for streaming scenarios, as the model needs to attend the full sequence. With the large memory consumption and long inference time, it is a challenge to apply such SSL frameworks to resource and latency-sensitive applications \cite{panchapagesan2021efficient}.

Knowledge distillation(KD) is a paradigm of transferring knowledge from a teacher model to a student model. It is a practical method to make the SSL-based models smaller or streaming. However, there is limited exploration about how to get a small and low latency model for online speech recognition tasks.
 
In this paper, we propose a flexible and efficient KD approach under the framework of W2V2, which can distill a large non-streaming model to a small streaming model (DistillW2V2) by taking advantage of huge untranscribed data. Our distillation method contains two steps. The first step performs KD to get a non-streaming student model. Then, a streaming student model is initialized from the non-streaming student model and the second KD is performed. To alleviate the accuracy degradation caused by the distillation process, we try to share as many model parameters as possible between teachers and students. An additional benefit of our two-stage distillation process is that it only needs to re-run the second step for different latency demands while reusing the previous pre-training, fine-tuning, and first distillation step. 

In terms of distillation objectives, to enable the student to learn output as well as intermediate knowledge from the teacher, we propose a distillation objective consisting of two parts, one is defined at the prediction layer and the other is defined at the hidden layers. We adopt an MSE distillation objective similar to \cite{jiao2019tinybert} for the hidden layers. For the prediction layer, to encourage student to learn sequence-level knowledge from teacher, we employ a LF-MMI \cite{povey2005discriminative,vesely2013sequence,povey2016purely,hadian2018end} based distillation objective, which has several successful applications in semi-supervised learning \cite{manohar2018semi,sheikh2020semi}. We modify the original method by directly generating denominator fst from massive text corpora rather than only using transcripts of labeled audio, which may be not enough to cover all possible sequences of the large unlabeled audio data.

To verify the effectiveness of our method, experiments are conducted on Gigaspeech, Librispeech, and an in-house dataset. For each proposed key point of our method, we also present an ablation study and analysis to demonstrate the effectiveness. Compared to the teacher model, our smallest streaming student is 8x faster. The relative degradation of WER gets 9\% on gigaspeech test set, 9.6\% on in-house test set, 13.6\%/23.4\% on Librispeech test-clean/other set, which is competitive to the similar topic works \cite{peng2021shrinking,dawalatabad2022two,yang2021knowledge,cao2021improving}.

\section{Related Work}

Recently, there are some KD methods aiming to reduce the large memory consumption and long inference time. For example, the recent study \cite{peng2021shrinking} explored the compression method for W2V2. The LightHuBERT \cite{wang2022lighthubert}, a once-for-all Transformer compression framework achieved comparable performance to the teacher model with a reduction of 29\% parameters. DistilHuBERT \cite{chang2022distilhubert} proposed to distill hidden representations from a HuBERT model directly and used a multi-task learning objective to encourage the transformer encoder to produce compact representations for multiple prediction heads. This method reduced HuBERT’s size by 75\%. FitHuBERT \cite{lee2022fithubert} explored a strategy of applying KD directly to the pre-trained teacher model, which reduced the model to 23.8\% in size and 35.9\% in inference time compared to HuBERT. Although the above methods have achieved a good model compression ratio, there is a lack of research on streaming ASR models.

The paper \cite{cao2021improving} first tried to construct a streaming W2V2 system by initializing the streaming student from the non-streaming teacher. However, the compression of model size was not explored in this work. The work \cite{yang2021knowledge} carried out the KD from a large pre-trained transducer model based  W2V2 to a much smaller streaming conformer model with a limited amount of labeled data. However, the performance is not ideal on the public test set. 

In this work, we introduce a two-stage distillation method to train a small and streaming student model,  which maintains good performance for online ASR applications.

\section{wav2vec 2.0}

As \cite{baevski2020wav2vec}, our baseline W2V2 architecture has three parts. Firstly, a multi-layer convolutional feature encoder $f: \mathcal{X} \rightarrow \mathcal{Z}$ is used to map input raw audio $\mathcal{X}$ to the latent speech representation. Secondly, a Transformer $g: \mathcal{Z} \rightarrow \mathcal{C}$ is adopted to build contextualized representation. Finally,  a quantization module $ \mathcal{Z} \rightarrow \mathcal{Q}$ is used to discretize the output of the feature encoder. 

The W2V2 model is first pre-trained with massive unlabeled data and then fine-tuned with limited labeled data. For the fine-tuning part, a linear prediction layer is added to the top of the pre-trained transformer and the model is usually optimized by minimizing a Connectionist Temporal Classification (CTC) \cite{graves2006connectionist} loss. \cite{vyas2021comparing} shows that LF-MMI and CTC can achieve similar results. In this work, we choose to use LF-MMI to keep consistent with our production system.
 
\section{Knowledge Distillation}

In this section, we will introduce our proposed KD process and MMI-based sequence-level distillation objective. 

\begin{figure}[htb]
  \centering
  \includegraphics[width=\linewidth]{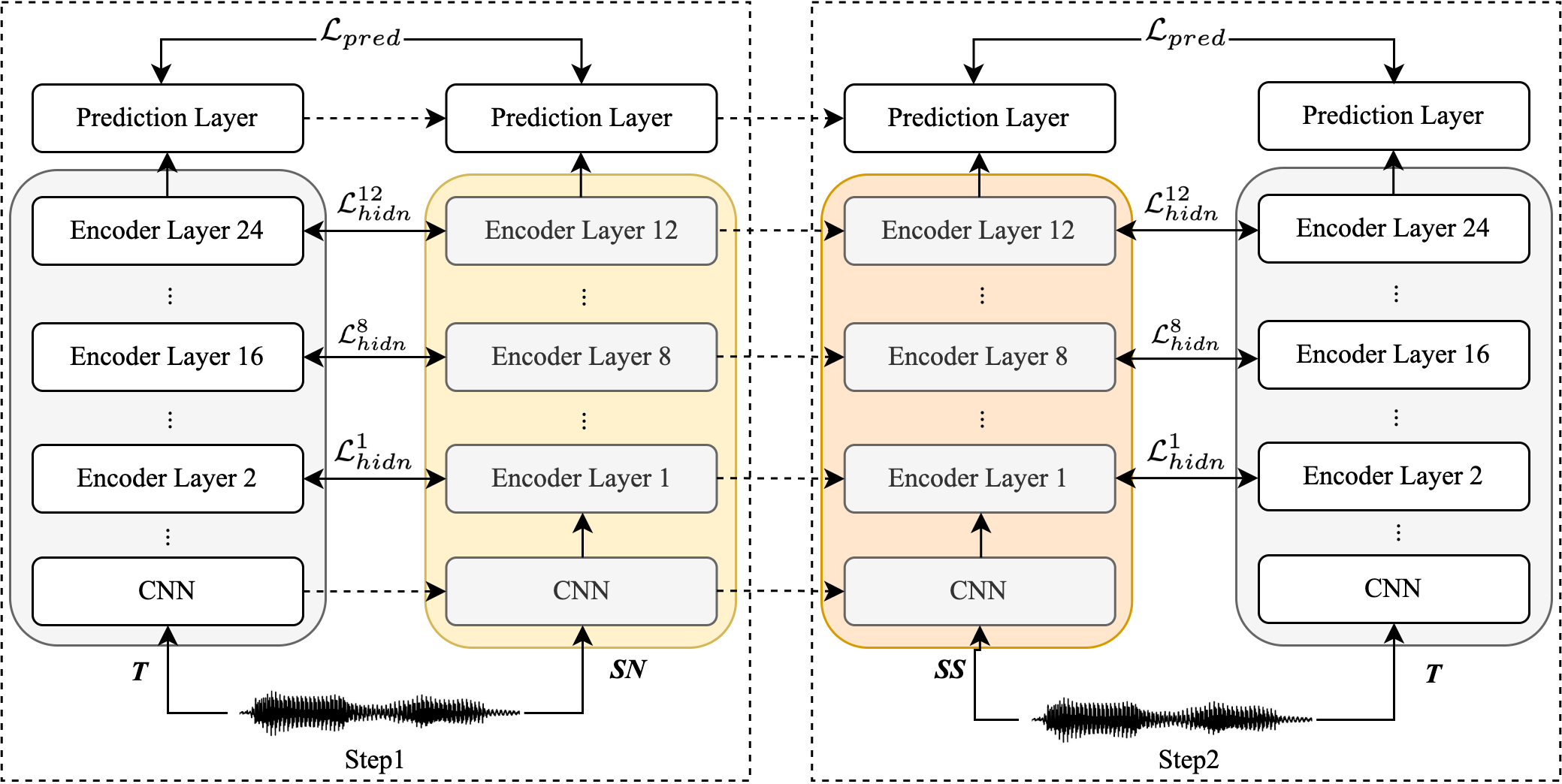}
  \caption{Our proposed two-step distillation process. The $\dashrightarrow$ indicates parameter initialization.}
  \label{fig:distilstage}
\end{figure}

\subsection{Distillation Process}

Figure~\ref{fig:distilstage} shows our two-step distillation process in which ${T}$ stands for the big and non-streaming teacher model, ${SN}$ means the small and non-streaming student model, ${SS}$ indicates the small and streaming student model. Similar to W2V2, we perform pre-training on massive unlabeled data and fine-tuning on limited transcribed data with LF-MMI criterion to get our teacher model ${T}$. The two-step distillation process can be described below:

\textbf{${T} \rightarrow {SN}$} In this step, we share the parameters of CNN encoder and prediction layer between ${T}$ and ${SN}$. We compress the teacher model's parameters such as the number of blocks, the dimension of the transformer encoder. Similar to \cite{cao2021improving}, we adopt group norm and a causal convolutional layer for ${SN}$. The distillation objective will be introduced in the next section.

\textbf{${SN} \rightarrow {SS}$} We follow the design of the streaming transformer as \cite{chen2021developing}, a chunk-transformer based architecture that combines Transformer-XL and chunk-wise processing. We initialize ${SS}$ with ${SN}$ and perform distillation again.

\subsection{Distillation Objective}
We perform knowledge distillation on hidden layers $\mathcal{L}_{hidn}$ and the prediction layer $\mathcal{L}_{pred}$, as different layers contain different acoustic properties \cite{baevski2021unsupervised}. The objective of knowledge distillation can be formulated as below.

\begin{footnotesize}
\begin{equation}
    \mathcal{L}_{distill} = (1-\alpha)\mathcal{L}_{hidn}+\alpha\mathcal{L}_{pred}\label{distill}\\
\end{equation}
\end{footnotesize}


\begin{footnotesize}
\begin{equation}
    \mathcal{L}_{hidn}=\sum_{i}MSE(\mathbf{H}^S_{i} \mathbf{W}_i, \mathbf{H}^T_{g(i)})\\
\end{equation}
\end{footnotesize}

\begin{footnotesize}
\begin{equation}
    \mathcal{L}_{pred}=\beta{MMI_{distill}}+(1-\beta)MSE(\mathbf{O^S},\mathbf{O^T}) \\
\end{equation}
\end{footnotesize}

 where $\mathbf{H}^S_{i}\in{R}^{l\times{d}}$ and $\mathbf{H}^T_{g(i)}\in{R}^{l\times{d'}}$ refer to the intermediate representation of the student model and the teacher model respectively, in which $d$ and $d'$ refer to the dimension of intermediate representation, $l$ indicates the sequence length. 
 $g(i)$ is a mapping function, which means that the $i$-th layer of the student model learns the information from the $g(i)$-th layer of the teacher model. 
 The $\mathbf{O}^S\in{R}^{l\times{m}}$ and $\mathbf{O}^T\in{R}^{l\times{m}}$ refer to the output of prediction layer, where $m$ is the dimension of output units. The trainable matrice $\mathbf{W}_i\in{R}^{d\times{d'}}$ is a linear transformation to match the dimension of teacher and student. $MMI_{distill}(.)$ indicates our proposed sequence-level KD objective. $\alpha$ is the weight of the prediction objective. $\beta$ is the weight of the sequence-level KD objective.

\textbf{LF-MMI based distillation Objective} The MMI criterion is used to discriminate the correct hypothesis from all hypotheses by maximizing the ratio as follows:

\begin{footnotesize}
\begin{equation}
    \mathcal{L}_{MMI}=\log{\frac{P(\mathbf{X}|W)P(W)}{\sum_{\overline{W}}P(\mathbf{X}|\overline{W})P(\overline{W})}} \\
\end{equation}
\end{footnotesize}

where $\mathbf{X}$, $W$ and ${\overline{W}}$ represent the acoustic feature, transcripts, and any possible hypothesis respectively. Similar to \cite{manohar2018semi}, we use the decoded lattices generated by the teacher model as pseudo supervisions for the student model. The LF-MMI distillation objective can be defined as:

\begin{footnotesize}
\begin{flalign}
MMI_{distill}=\log{\frac{\sum_{W\in{G^T_{num}}}P_{S}(\mathbf{X}|W)P(W)}{\sum_{\overline{W}\in{G_{den}}}P_{S}(\mathbf{X}|\overline{W})P(\overline{W})}}&\label{distil_ori} 
\end{flalign}
\end{footnotesize}

where $P_S(.)$ is the likelihood generated by the student model, $G^T_{num}$ is the numerator graph generated from the decoded lattice of the teacher model, $G_{den}$ is the denominator graph which approximately represents all possible state sequences. 
Different from \cite{manohar2018semi}, whose $G_{den}$ is generated from the phone sequence of labeled audio by force-alignment, we generate $G_{den}$ from the large raw text corpora to enrich the diversity of hypotheses. For the pronunciation of each word, we just randomly choose one from the lexicon, and also randomly insert silence phone between words and at sentences' boundaries similar to \cite{hadian2018end}.

\section{Experiment}

\subsection{Experiment setup}

\subsubsection{Dataset}
To evaluate our method, we perform experiments on Gigaspeech \cite{chen2021gigaspeech},  Librispeech \cite{panayotov2015librispeech}, and an in-house corpus. For Gigaspeech, we use the XL subset (10,000h) as unlabeled data and the S subset (250h) as labeled data. For Librispeech, we use LV-60k (5,3200h) in \cite{baevski2020wav2vec} as unlabeled data, and train-clean-100 (100h) as labeled data. For the in-house corpus, we use 38,000h of unlabeled data, and 300h of labeled data which consist of non-native English spoken by Chinese with personally identifiable information removed. Notably, we used the unlabeled data in the pre-training and distillation stages.

\subsubsection{Model architecture}

Table~\ref{tab:sizeconf} describes the details of models. For the model configuration, we mainly consider the channels of the first two layers of the CNN encoder, the dimension and the block number of the transformer model. We evaluate the real-time factor (RTF) on CPU (Intel 8255C) with 32GB RAM using a single thread. The layer mapping function adopted here is $g(i)=2i$. For S4 and S5, we just remove the 4th and 8th layers’ mapping. As shown in the table, the smallest student (S5) is 8x faster and 12x smaller than the teacher model (T1).

\begin{table}[pth]
   \caption{Model configuration. CNN-Enc.{$^{1-2}$} indicates the first two layers in the CNN feature encoder.}
  \label{tab:sizeconf}
  \centering
  \begin{adjustbox}{max width=\linewidth}
  \begin{tabular}{ lccccccc }
    \toprule
     \multicolumn{1}{c}{\multirow{2}{*}{Model}} & \multirow{2}{*}{CNN-Enc.{$^{1-2}$}}  & \multirow{2}{*}{Encoder} & \multirow{2}{*}{FFN} &\multirow{2}{*}{Blocks} &\multirow{2}{*}{Param}   &\multirow{2}{*}{RTF}\\\specialrule{0em}{2pt}{2pt}
     &dim &dim &dim & &  & \\
    \midrule
    \textbf {teacher} \\
    $\quad$T1 & 512 & 1024 & 4096 & 24 & 317M &0.889 \\
    \midrule
    \textbf {student} \\
    $\quad$S1 & 512 & 768 & 3072 & 12 & 95M &0.357 \\
    $\quad$S2 & 512 & 512 & 2048 & 12 & 44M &0.196 \\
    $\quad$S3 & 512 & 384 & 1536 & 12 & 27M &0.135 \\
    $\quad$S4 & 512 & 384 & 1536 & 10 & 23M &0.121 \\
    $\quad$S5 & 256 & 384 & 1536 & 10 & 22M &0.111 \\
    \bottomrule
  \end{tabular}
  \end{adjustbox}
\end{table}

\begin{table}[th]
  \caption{WER results of the knowledge distillation process. (600,48) indicates that the chunk size is 48 frames, and the history chunk size is 600 frames. $(+\infty,+\infty)$ indicates a full context chunk size.}
  \label{tab:Streamingdistill}
  \centering
  \begin{adjustbox}{max width=\linewidth}
  \begin{tabular}{ ccccccc }
    \toprule
    \multicolumn{1}{c}{\multirow{2}{*}{Steps}} &\multirow{2}{*}{(hist, chunk)} &\multirow{2}{*}{latency} & \multicolumn{2}{c}{test} & \\ \cmidrule(lr){4-5} & & &gigaspeech & in-house \\
    \midrule
    $\quad$Teacher & $(+\infty,+\infty)$ & - & 11.84 &17.56\\
    \midrule
    $\quad$Single step & (600,48)& 480ms &15.58& 22.36  \\
    \midrule
    \textbf{this work} \\
    $\quad$First step & $(+\infty,+\infty)$ & - &12.37& 18.41  \\
    $\quad$Second step & (600,48)& 480ms & 12.90 & 19.25\\
    \bottomrule
  \end{tabular}
  \end{adjustbox}
\end{table}

\subsubsection{Training details}
 
Our training is mainly based on fairseq\footnote{https://github.com/pytorch/fairseq}. 
We follow \cite{baevski2020wav2vec} to do pre-training and fine-tuning to get the teacher model.
For the distillation part, we follow the standard Kaldi recipe of train-clean-100 and gigaspeech\_train\_s  to generate the tree and transition model. 
Models are trained on 16 V100 GPUs with Adam and a tri-state learning rate schedule where the learning rate is warmed up for the first 10\% of updates, held constant for the next 40\%, and then linearly decayed to a final scale of 0.05 for the remainder.
Epochs and peak learning rates for the two-step distillation are 30/15, and 5e-4/1e-4. 
Volume and pitch augmentation is applied on raw audio with a probability of 0.5 and the scale range is 0.3-3.0 for volume and the shift range is 0.9-1.1 for pitch.

\subsection{Results of two-step distillation} 

Our experiments are evaluated on the Gigaspeech and the in-house corpus. 
For the first step, T1 and S5 of Table~\ref{tab:sizeconf} are used. For the parameters of the CNN feature encoder, we just share part of the weight without the first two layers.
For the second step, the chunk-transformer is adopted to make the student model streaming.
The sizes of history chunks and the current chunk are 600 and 48 frames, indicating the average look-ahead time is 480ms.

The results are shown in Table~\ref{tab:Streamingdistill}.
Compared to the single-step process, our two-step distillation achieves a relative WER improvement of 15\%. Since two variables of model compression and streaming existing in the entire process, the single-step student's parameters have to be randomly initialized.
The performance of the streaming model would be further improved by initializing with a non-streaming model before distilling.

\subsection{Ablation Study}

\subsubsection{Training objective}
The training object plays a vital role during knowledge distillation.
In this part, we will validate the effectiveness of our proposed training objective defined by Eq.\eqref{distill}.
Experiments are focused on the first step of the distillation process and results are shown in Table~\ref{tab:distillMethod}.

First of all, we investigate the effect of different criteria for the prediction layer.
Results show that LF-MMI loss (M2) is better than MSE (M1).
Then we try to add $\mathcal{L}_{hidn}$ to get model M3, which shows that the hidden layer distillation has a positive effect on the accuracy of the student model. Furthermore, we combine MSE and LF-MMI objectives for $\mathcal{L}_{pred}$ to get model M4, which is slightly better than M3 for in-house test data.

\begin{table}[th]
  \caption{WER results of different training objectives for the first step distillation.}
  \label{tab:distillMethod}
  \centering
  \begin{adjustbox}{max width=1.0\linewidth}
  \begin{tabular}{ lccccc }
    \toprule
     \multicolumn{1}{c}{\multirow{2}{*}{Objective}}  &\multirow{2}{*}{hidden} &\multirow{2}{*}{pred} & \multicolumn{2}{c}{test}  & \\  \cmidrule(lr){4-5} & & & Gigaspeech & in-house \\
    \midrule
    \textbf {Teacher} \\
    $\quad$T1 & - & - & 11.84 & 17.56 \\
    \midrule
    \textbf {Student} \\
    $\quad$M1($\alpha=1,\beta=0$) & - & MSE  & 14.29 & 20.83  \\
    $\quad$M2($\alpha=1,\beta=1$) & - & LF-MMI  & 13.54 & 20.22  \\
    $\quad$M3($\alpha=0.8,\beta=1$) & MSE & LF-MMI & 12.35 & 18.60  \\
    $\quad$M4($\alpha=0.8,\beta=0.8$)& MSE & MSE/LF-MMI & 12.37 & 18.41 \\
    \bottomrule
  \end{tabular}
  \end{adjustbox}
\end{table}

\subsubsection{Effect of compression ratio}
In this section, we study the effect of compression ratio on the performance of streaming student models which have the same streaming setup. The size of the models can be found in Table~\ref{tab:sizeconf}. The result is shown in Table~\ref{tab:compress-scale1}. 
As the compression ratio increases, the performance of the student model deteriorates slightly.
However, compared with the 95M model (S1), the 22M model (S5) has a 4.3x compression rate with only 4.0\%-4.4\% relative WER degradation.
Comparing S5 and S4, there is only a small difference in WER, indicating that our distillation method is also effective for the CNN encoder.


\begin{table}[th]
  \caption{WER results of streaming models with different model sizes.}
  \label{tab:compress-scale1}
  \centering
  \begin{adjustbox}{max width=1\linewidth}
  \begin{tabular}{ lccccccc }
    \toprule
     \multicolumn{1}{c}{\multirow{2}{*}{Test}}& \multirow{2}{*}{T1}& \multirow{2}{*}{S1}& \multirow{2}{*}{S2} &  \multirow{2}{*}{S3} &  \multirow{2}{*}{S4} & \multirow{2}{*}{S5} \\\specialrule{0em}{2pt}{2pt}
    \midrule
    $\quad$gigaspeech & 11.84 & 12.35 & 12.66 & 12.88 & 12.91  & 12.90\\\specialrule{0em}{1pt}{1pt}
    $\quad$in-house & 17.56 &18.51 & 18.83 & 19.07  &19.15 & 19.25  \\

    \bottomrule
  \end{tabular}
  \end{adjustbox}
\end{table}

\subsection{Comparison with other works}

As shown in Table~\ref{tab:finalresult}, we compare our method with several existing works on the widely used Librispeech data set. The student model corresponds to S5 of Table~\ref{tab:sizeconf}. The streaming configuration is the same as in Sec4.2. For decoding, we use the 4-gram and Transformer language model(LM) described in\cite{baevski2020wav2vec}. Compared with other KD methods, with a relatively high compression ratio, our streaming model achieves very competitive results. To be mentioned, we use LV-60k as unlabeled data, while others use LS-960. We leave fair comparison as our future work.

\begin{table}[th]
  \caption{WER results on Librispeech.}
  \label{tab:finalresult}
  \centering
  \begin{adjustbox}{max width=1\linewidth}
  \begin{tabular}{ lcccccccc }
    \toprule
     \multicolumn{1}{c}{\multirow{2}{*}{System}} &\multirow{2}{*}{LM}& \multirow{2}{*}{Streaming/latency} &\multirow{2}{*}{labeled data} &\multirow{2}{*}{model size} &\multicolumn{2}{c}{Librilight} & \\ \cmidrule(lr){6-7} & & & (hour) &(m) & test-clean & test-other\\
    \midrule
    \textbf{W2V2 transducer\cite{yang2021knowledge}} \\
     $\quad$teacher & - & no & 100h & 99.2M & 5.2  & 11.8 \\
     $\quad$student & - & yes/280ms & 100h & 9.7M & 9.8  & 18.2 \\
    \textbf{Streaming W2V2\cite{cao2021improving}} \\
     $\quad$teacher &4-gram & no & 100h & 95M & 3.3  & 8.1 \\
     $\quad$student &4-gram &yes/480ms & 100h & 95M & 3.5  & 8.7 \\
     \textbf{Compressing W2V2 \cite{peng2021shrinking}} \\
     $\quad$teacher &Transformer &no & 100h & 317M & 2.63  & - \\
     $\quad$student &Transformer &no & 100h & 166M & 6.6  & - \\
    \midrule
     \textbf{this work}\\
    $\quad$teacher &4-gram &no  & 100h & 317M &2.91  &5.43 \\
    $\quad$Student in Step1 &4-gram & no  & 100h  & 22M  &3.29 &6.63 \\
    $\quad$Student in Step2 &4-gram & yes/480ms & 100h  & 22M &3.54  &7.57\\
    \textbf{this work} \\
    $\quad$teacher &Transformer & no  & 100h & 317M & 2.2  & 4.26 \\
    $\quad$Student in Step1 &Transformer &no & 100h  & 22M  & 2.42 & 4.83 \\
    $\quad$Student in Step2 &Transformer & yes/480ms & 100h  & 22M &2.50  &5.27\\
    \bottomrule
  \end{tabular}
  \end{adjustbox}
\end{table}

\section{Conclusion}

In this work, we propose a knowledge distillation method for model compression and streaming. We explore both the MSE-based objective and the proposed modified LF-MMI objective. The ablation study shows that intermediate knowledge is important for the distillation results, and the LF-MMI criterion is better than the MSE criterion for the prediction layer. Our small-footprint streaming model (DISTILLW2V2) gets competitive WERs by using only a small subset of labeled data. In future work, we intend to explore more sequence-level objectives for distilling streaming models.

\bibliographystyle{IEEEbib}
\bibliography{strings,refs}

\end{document}